\documentclass[journal,twoside, a4paper]{IEEEtran}
\IEEEoverridecommandlockouts
\usepackage{cite}
\usepackage{amsmath,amssymb,amsfonts}
\usepackage{algorithmic}
\usepackage{graphicx}
\usepackage{textcomp}
\usepackage{xcolor}
\usepackage{hyperref}
\usepackage[ruled,vlined,noresetcount]{algorithm2e} 
\usepackage{amsmath}

\def\BibTeX{{\rm B\kern-.05em{\sc i\kern-.025em b}\kern-.08em
    T\kern-.1667em\lower.7ex\hbox{E}\kern-.125emX}}
\usepackage[a4paper, total={7in, 9in}]{geometry}
\begin{document}

\title{Deep Learning Autoencoders for Reducing PAPR in Coherent Optical Systems}

\author{\IEEEauthorblockN{Omar Alnaseri\IEEEauthorrefmark{1}, 
Ibtesam R. K. Al-Saedi\IEEEauthorrefmark{4}, 
Yassine Himeur\IEEEauthorrefmark{5} and Hongxiang Li\IEEEauthorrefmark{6}
}
\\

\IEEEauthorblockA{\IEEEauthorrefmark{1}Department of Electrical Engineering, DHBW University, Ravensburg, Germany, \\ Email: alnaseri.omar@dozent.dhbw-ravensburg.de}\\
\IEEEauthorblockA{\IEEEauthorrefmark{4}Department of Communication Engineering, University of Technology - Iraq, Baghdad, Iraq}\\
\IEEEauthorblockA{\IEEEauthorrefmark{5}College of Engineering and Information Technology, University of Dubai, Dubai, United Arab Emirates}\\
\IEEEauthorblockA{\IEEEauthorrefmark{6}Department of Electrical and Computer Engineering, University of Louisville, Kentucky, United States}
}

\maketitle

\pagenumbering{gobble}

\begin{abstract}
This paper presents an innovative approach to mitigating the peak-to-average power ratio (PAPR). The proposed method uses a deep learning model called autoencoders (AEs) to simplify the process and avoid the complex calculations of traditional methods such as selective mapping (SLM). Unlike SLM, our approach does not need side information about the PAPR distribution. Through simulations of coherent optical orthogonal frequency division multiplexing (CO-OFDM) systems, the AE-based model offers substantial enhancements in both PAPR reduction and bit error rate (BER) performance when compared to conventional techniques. An error-free transmission can be acheived with a reduction in PAPR exceeding 10 dB compared to the original signal and a 1 dB advantage over SLM. In particular, the AE model achieves the best BER performance of \(2 \times 10^{-6}\) at 44 dB OSNR, surpassing traditional methods. Furthermore, the model demonstrates robustness against noise and nonlinear distortions, making it appropriate for optical channels experiencing diverse levels of impairment. This innovative technique has the potential to revolutionize next-generation optical communication systems by enabling efficient and reliable data transmission.
\end{abstract}


\section{Introduction}
Orthogonal frequency division multiplexing (OFDM) is widely preferred as a multi-carrier modulation format. Coherent optical orthogonal frequency division multiplexing (CO-OFDM) emerges as a promising technology for long-haul fiber optic communication. It is capable of effectively compensating for signal distortions induced by chromatic dispersion (CD) and polarization-mode dispersion (PMD), while simultaneously facilitating increased data transmission rates \cite{article0}. However, a significant limitation of CO-OFDM technology is its high PAPR, which has the potential to significantly affect transmission bit error rate (BER) performance \cite{huleihel2024low,zhang2024intelligent}. High PAPR can push both the amplifier and the IQ modulator into a nonlinear operating region, causing signal distortion. Recently, a transmission rate of 110 Gbps was achieved over a 105 km optical fiber distance using standard single mode fiber (SSMF) \cite{article1}. This was accomplished by using a 60 GHz radio-over-fiber (RoF) transmission system and a 16-quadrature amplitude modulation (16-QAM) OFDM baseband signal. In order to alleviate the effects of laser phase noise within a 16-QAM CO-OFDM system utilizing a 1024-point fast Fourier transform (FFT), a technique called subcarrier-index modulation OFDM (SIM-OFDM) was introduced in \cite{article2}. However, the high PAPR resulting from the combination of 16-QAM modulation and a large 1024-point FFT was not addressed, despite its tendency to induce considerable nonlinear distortions in the amplifiers, modulator and optical fiber \cite{lipovac2024otdr}. 

One of the main challenges in CO-OFDM systems lies in managing high PAPR, which increases with higher modulation formats and larger FFT sizes, as seen with 16-QAM and 1024-point FFTs \cite{nanthaamornphong2025phase}. High PAPR can result in severe nonlinear distortions when signals pass through optical components like amplifiers and modulators, pushing these into their nonlinear operating regions. Traditional PAPR reduction techniques, such as amplitude clipping and selective mapping (SLM), come with limitations \cite{article12}. Clipping, while simple, introduces unwanted noise, affecting the overall quality of the signal, while SLM avoids BER degradation but adds computational complexity due to multiple FFT operations and requires additional side information, complicating the design of the system \cite{article11}. Discrete Fourier transform spread (DFT-spread) effectively reduces PAPR and fiber nonlinearity by distributing data energy across subcarriers before IFFT, which can lower peak power\cite{jan2013experiment}. It bypasses complex searches and side information, but demands numerous DFT operations, which are computationally intensive. Addressing PAPR effectively without compromising BER performance and system simplicity motivates the exploration of machine learning, particularly deep learning(DL)-based approaches, to develop a robust PAPR reduction method that minimizes computational load and the need for side information \cite{kumar2024low}.

In contrast, the application of machine learning, with an emphasis on deep learning (DL), has been investigated in the domain of communication systems \cite{article3,article4, alnaseri2024review}. For example, in \cite{article5}, the notion of employing a deep neural network autoencoder (AE) designed for end-to-end learning in communication systems was expanded to encompass OFDM transmission over multipath channels. Inspired by \cite{article13, al2025leveraging}, we explore the use of AE encoder and decoder components in CO-OFDM systems. Through end-to-end training, the AE model learns to accurately reconstruct the input data at the receiver, even in the presence of noise channel impairments. The proposed AE significantly can avoid the SLM computational complexity by eliminating the need for additional IFFT operations and complex search algorithms. Furthermore, the approach eliminates the need to transmit any side information. Our results demonstrate that the AE approach achieves a substantial reduction in PAPR and improved BER performance. In general, the main contributions of the proposed study are as follows.

\begin{itemize}
\item The study presents a novel approach to reduce PAPR in CO-OFDM systems by using a DL-based AE model, which simplifies the process and reduces computational complexity compared to traditional methods.

\item Unlike conventional techniques such as SLM, which rely on side information, the proposed AE-based model does not require any additional side information for PAPR reduction. This facilitates implementation and diminishes the overhead in communication systems.

\item Through simulations, the AE-based model demonstrates a substantial reduction in PAPR (over 10 dB compared to the original signal and 1 dB over SLM) and improved BER performance, achieving error-free transmission in scenarios where traditional techniques would struggle.

\item The model is robust against noise and nonlinear distortions common in optical channels, which are often challenging to effectively handle traditional approaches. This makes the proposed method suitable for use in optical communication systems with varying channel impairments.

\item By enabling efficient and reliable data transmission in optical channels, the AE-based PAPR reduction approach has significant potential to advance next-generation optical communication systems, providing a scalable solution for high-performance, long-distance fiber optic networks.
\end{itemize}

\section{Peak-to-Average Power Ratio (PAPR)}
OFDM offers several advantages, but one of its key challenges is its high PAPR. This occurs because a time domain OFDM signal is the sum of multiple subcarriers, each modulated independently and with unique phases. As these subcarriers can add constructively or destructively, the resulting signal can have peak values significantly exceeding the average power of the signal. In particular, OFDM symbols with longer durations, which correspond to larger IFFT sizes, are more prone to high PAPR. Mathematically, the PAPR of a transmitted signal, denoted as $w(t)$, is defined as:
\begin{equation}
\label{eq:003}
PAPR_{w}(dB) = 10 \cdot \log_{10}\left(\frac{\max\left[w(t)^{2}\right]}{\text{mean}\left[w(t)^{2}\right]}\right).
\end{equation}

The high PAPR in OFDM signals poses two major challenges: (1) drive amplifiers and IQ modulator into nonlinear region; (2) optical fiber nonlinearity: The high peak power caused by PAPR can push the optical fiber  into their nonlinear regimes. This can lead to signal distortion and performance degradation. (3) Limited the digital-to-analog converter (DAC) resolution: PAPR requires a high-resolution DAC to faithfully reproduce the signal's peaks without clipping. However, DACs have limitations in their resolution, which can introduce additional distortion if the PAPR is too high.

\section{Clipping Technique}
Clipping is a straightforward technique for PAPR reduction that involves cutting signal amplitudes that exceed a predefined threshold. Although this approach is computationally efficient, it distorts both the signal components within the desired frequency band and those outside of it. Although out-of-band distortion can be mitigated through filtering, in-band distortion persists. The clipping threshold can be defined as:
$ T_{clip} \gets P_{clip} \cdot \max_{i} |x| $, where $P_{clip}$ is the clipping ratio, and $x$ represents the signal in the time domain. The clipped data $\hat x_{k}$ is determined by:
\begin{equation}
\label{eq:9}
\hat x_{k} = \begin{cases}
T_{clip} \cdot e^{j \angle(x_{k})} \ \ \ \text{if } |x_{k}| > T_{clip} \\
x_{k} \ \ \ \ \ \ \ \ \ \ \ \ \ \ \ \text{otherwise}
\end{cases}
\end{equation}
where $k$ is the time-domain signal index, $\angle(x_{k})$ represents the phase of $k$-th data in radians, to maintain the original phase. 

\section{Selective Mapping (SLM)}
\begin{figure}[htbp]
    \centering
    \includegraphics[width=1.0\linewidth]{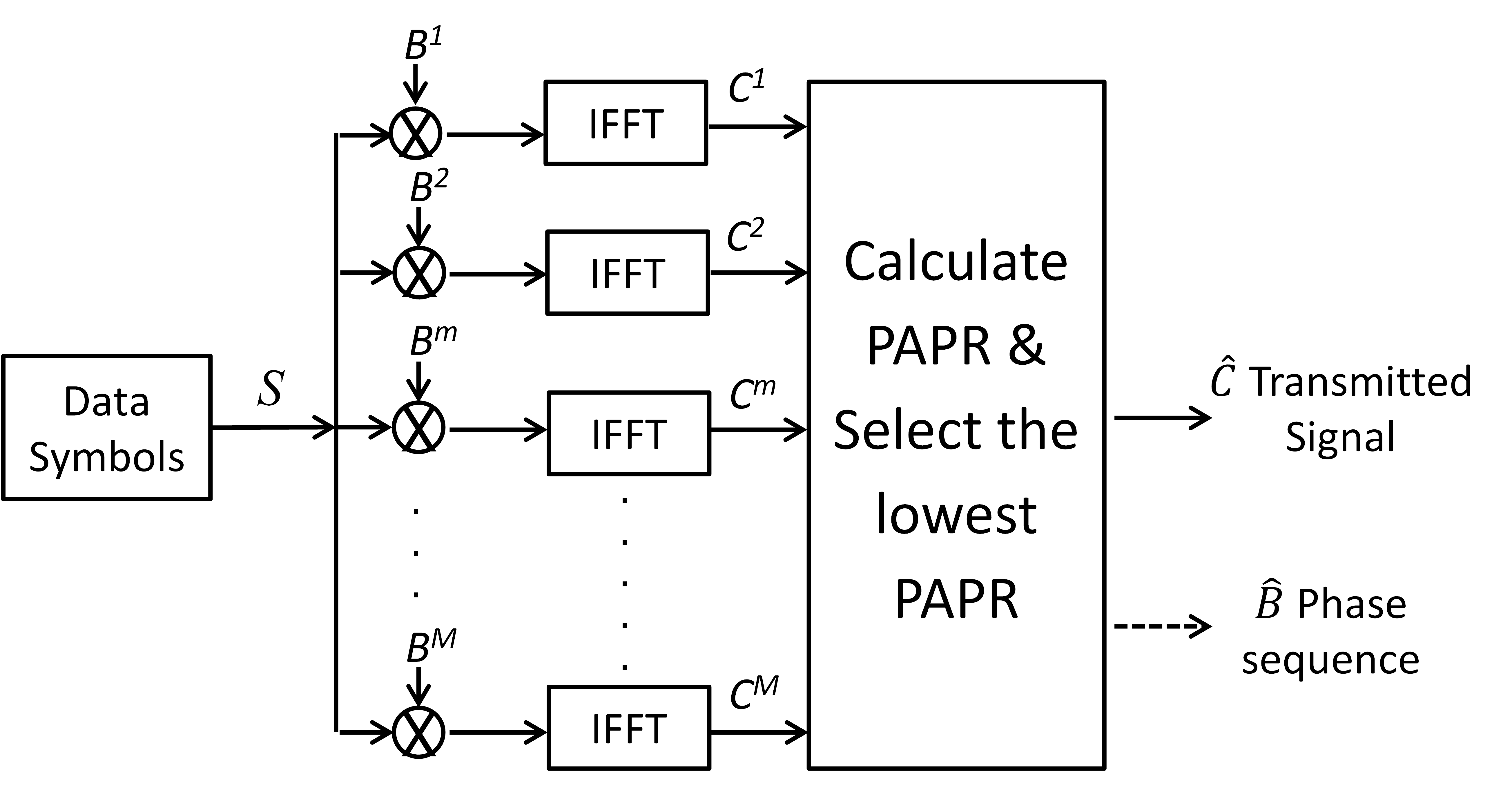}
    \caption{A block diagram of SLM technique}
    \label{fig1_1}
\end{figure}

Selective Mapping (SLM) is an alternative PAPR reduction technique that preserves signal integrity. As shown in Fig.~\ref{fig1_1}, it generates multiple OFDM symbol candidates by applying several phase sequences. The one with the lowest PAPR is selected for transmission as,  
\begin{equation}
\label{eq:8}
\widehat{C} = \underset{1\leq m\leq M}{argmin} \left ( \frac{\max\limits_{1\leq n\leq N} [|C_m|^{2}]}{ave\left [ |C_m|^{2} \right ]} \right),
\end{equation}
where $max[]$ and $ave[]$ represent the maximum and average of OFDM symbol power values of $N$ subcarriers. To enable correct signal recovery at the receiver, the applied phase sequence must be conveyed as side information. However, the implementation of the SLM requires multiple IFFT operations, increasing the complexity of the system.

\section{End-to-end AE Learning}
\begin{figure*}[htbp]
    \centering
    \includegraphics[width=0.75 \linewidth]{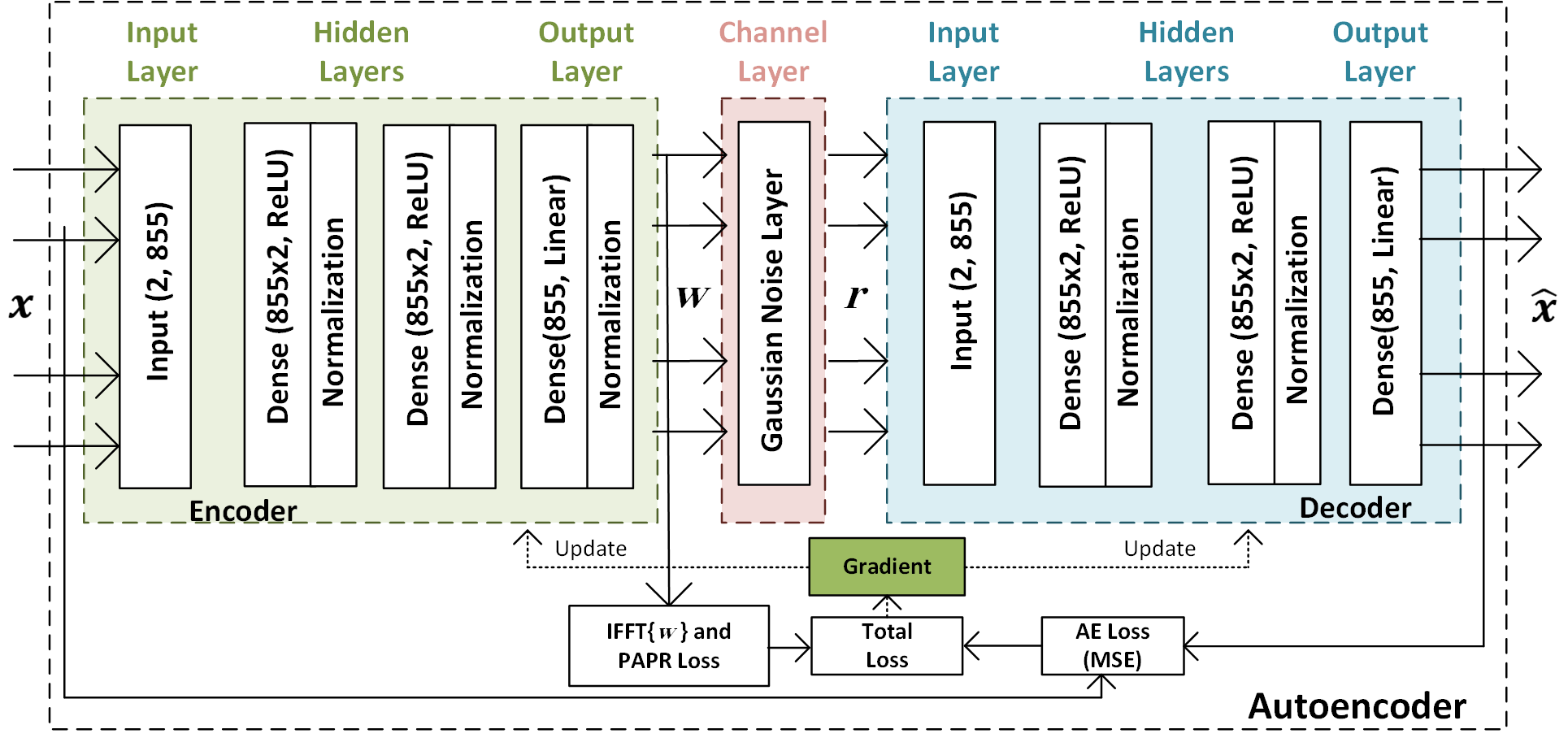}
    \caption{AE structure with 2 hidden layers in both the transmitter and receiver, along with Gaussian noise layer as a channel layer}
    \label{fig2}
\end{figure*}

An AE architecture is proposed to generate robust data ($w$) with minimal PAPR as shown in Fig.~\ref{fig2}. It consists of an input layer with dimensions (2, 855), followed by 2 dense hidden layers (2x855 dimensions) with rectified linear unit (ReLU) activation functions in the encoder part. The encoder output layer is a linear activation layer with 855 dimensions. A channel layer incorporating Gaussian noise with varying standard deviations ($\sigma$ = 0.1, 0.16, 0.2, and 0.35) is investigated. The decoder mirrors the encoder structure, with two dense hidden layers with ReLU activations and a final linear output layer.

The proposed end-to-end learning utilizes two loss functions to optimize the AE for mitigating PAPR in coherent communication systems. The first loss function, denoted AE loss in Eq.~(\ref{eq:1}), measures the mean squared error between the input QAM symbols ($x$) and the reconstructed symbols of the decoder ($\hat{x}$). This loss function ensures that the AE faithfully reproduces the original signal. 

\begin{equation}
\label{eq:1}
Loss_{ae}(\hat{x}, x) = \frac{1}{n} \sum_{i=1}^{n} (x_{i} - \hat{x}_{i})^2, 
\end{equation}
where $n$ is the number of QAM symbols, $x$ is the QAM input symbols to the AE, and $\hat{x}$ is the reconstructed output symbols of the AE. For the second loss function, the AE encoder output ($w$) is first converted to time-domain representation as, 
\begin{equation}
\label{eq:2}
\text{IFFT} \left\{ w_{i} \right\} = \hat{w}_{k}
\end{equation}
where $w_{i}$ is the $i$-th element of the frequency-domain representation of the generated latent representations (i.e. encoder output), and $\hat{w}$ is the $k$-th element of the time-domain representation of the generated latent representations. Then the PAPR loss in Eq.~(\ref{eq:3}), is introduced to address the PAPR issue, and aims to minimize the PAPR of the generated latent representations.
\begin{equation}
\label{eq:3}
Loss_{papr}(w) = 10 \log_{10}\left(\frac{\max\left(\sqrt{\sum_{k=1}^{n} \hat{w}_{k}^2}\right)}{\frac{1}{n} \sum_{k=1}^{n} \sqrt{\sum_{k=1}^{n} \hat{w}_{k}^2}}\right),
\end{equation}

\section{Simulation System Setup}
\begin{figure*}[t!]
    \centering
    \includegraphics[width=0.78\linewidth]{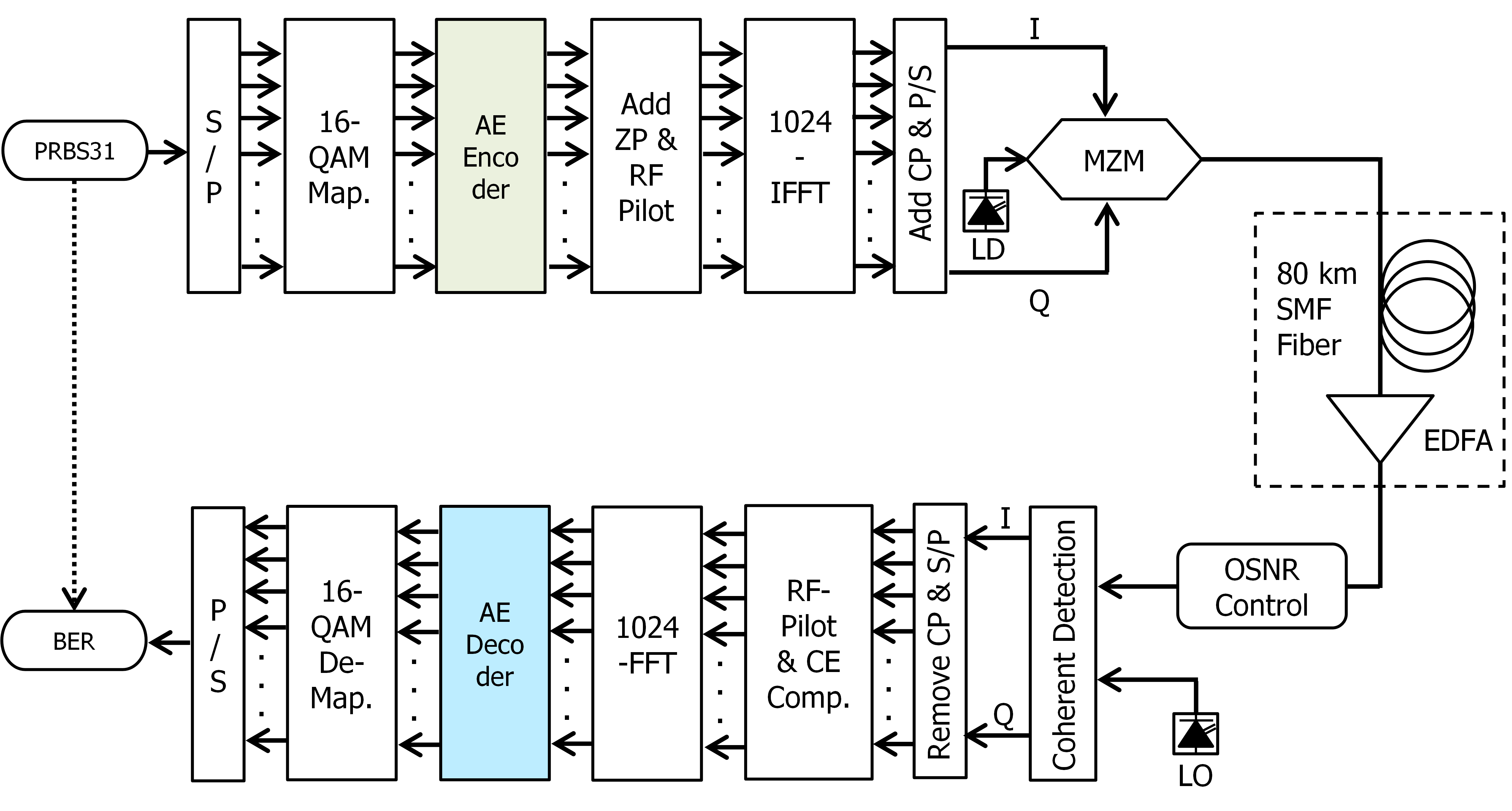}
    \caption{CO-OFDM system transmission based on AE}
    \label{fig3}
\end{figure*}

The setup for the CO-OFDM system with AE parts is shown in Fig.~\ref{fig3}. The OFDM communication system, which includes both the encoder and the decoder of the AE (depicted in light green and light blue, respectively), was implemented using Python. These two parts have been trained to generate resilient latent representations and reconstruct original 16-QAM symbols while considering varying standard deviation of Gaussian noise in the channel layer.

At the transmitter, the first step is mapping a 31-bit Pseudo-Random Binary Sequence (PRBS) to a 16-QAM constellation. The AE encoder then processes this signal, producing 855 symbols. To prevent aliasing, 128 zero-padded subcarriers are added to the encoder output. In addition, a 20 subcarrier guard band is included on either side of the RF-pilot, as described in \cite{article7} and \cite{article8}. The RF-pilot is used to compensate for laser phase noise. Then, the frequency domain signal is transformed into the time domain using a 1024-point IFFT. A 12.5\% cyclic prefix is added to mitigate Inter-Symbol Interference (ISI) caused by chromatic dispersion. OFDM signals in the baseband are then encoded onto the optical carrier using a Mach-Zehnder modulator for optical IQ modulation, utilizing a laser with a 100 kHz linewidth. To specifically evaluate the effect of PAPR, the optical link is simulated using Python \cite{article9}, consisting of 10 spans. Each span covers a distance of 80 km using Standard Single-Mode Fiber (SSMF), and an erbium-doped fiber amplifier (EDFA) with a gain of 16 dB is used to compensate for the 0.2 dB/km fiber loss experienced in each span. The fiber chromatic dispersion and nonlinearity parameter are set to 16 ps/nm/km, and 1.3 $\text{(W.km)}^{-1}$ respectively, with 2.5 Gb/s baudrate. The launch power is set to 0 dBm.

On the coherent receiver side, an optical hybrid $90^{\circ}$ is utilized to detect the received optical IQ signal with a 100 kHz laser linewidth as a local oscillator, along with balanced detectors that convert optical signals to electrical signals \cite{article9}. Afterwards, the OFDM decoder processes the electrical signal. Within the OFDM decoder, first CP removal is performed, then the RF-pilot is filtered out, and then multiplied by the conjugate of the received signals to mitigate the phase noise in time domain. The characteristic of the communication channel transfer function is affected by the chromatic dispersion in the fiber, causing a frequency-dependent phase shift, referred to as the phase disparity caused by CD among subcarriers \cite{article10}. In order to address this phase discrepancy, each OFDM frame comprises 52 OFDM symbols, including four training symbols, which provide sufficient information for precise channel estimation and correction. The AE decoder that has been trained subsequently rebuilds the original 16-QAM symbols. The QAM demapper converts these symbols into a bit sequence, which is then contrasted with the original 31-bit PRBS to evaluate the BER.

The algorithm \ref{algo1} outlines the proposed DL-based AE method to reduce PAPR in coherent optical OFDM systems. Starting with 16-QAM symbol generation from binary sequences, the symbols are encoded through dense layers optimized to lower PAPR. Gaussian noise simulates channel conditions, and a decoder reconstructs the symbols. Two loss functions are used during training: reconstruction loss (to ensure accurate symbol recovery) and PAPR loss (to penalize high PAPR). These are minimized together to improve both signal quality and PAPR. The AE is trained iteratively until convergence, and its performance is evaluated through BER and Complementary Cumulative Distribution Function (CCDF) metrics, showing it as a robust alternative to traditional PAPR reduction methods like SLM.

\begin{algorithm}[t!]
\caption{Proposed AE Methodology for Reducing PAPR in CO-OFDM Systems}
\label{algo1}
\SetAlgoLined
\textbf{Input:} 16-QAM symbol sequence $\mathbf{x}$, Gaussian noise standard deviation $\sigma$, maximum iterations $N$ \\
\textbf{Output:} Reconstructed symbol sequence $\hat{\mathbf{x}}$, PAPR-reduced latent representation $\mathbf{w}$ \\
\textbf{Step 1: Data Preparation and Mapping} \\
Generate PRBS bits, map to 16-QAM symbols $\mathbf{x}$, and normalize. \\
\textbf{Step 2: Encoder Network} \\
Input $\mathbf{x}$ to encoder; pass through dense layers with ReLU to obtain latent variables $\mathbf{w}$ optimized for lower PAPR. \\
\textbf{Step 3: Channel Layer} \\
Add Gaussian noise to $\mathbf{w}$, producing $\mathbf{w}_{noisy} = \mathbf{w} + \mathcal{N}(0, \sigma)$. \\
\textbf{Step 4: Decoder Network} \\
Decode $\mathbf{w}_{noisy}$ to reconstruct symbols $\hat{\mathbf{x}}$. \\
\textbf{Step 5: Loss Calculation} \\
Compute Mean Squared Error (MSE) loss for AE reconstruction:
\[
Loss_{ae} = \frac{1}{N} \sum_{i=1}^{N} (\mathbf{x}_i - \hat{\mathbf{x}}_i)^2
\]
Compute PAPR loss from $\mathbf{w}$'s time-domain IFFT:
\[
Loss_{papr} = 10 \log_{10}\left(\frac{\max(|\text{IFFT}(\mathbf{w})|^2)}{\mathrm{mean}(|\text{IFFT}(\mathbf{w})|^2)}\right)
\]
Minimize combined loss $Loss_{total} = Loss_{ae} + \lambda \cdot Loss_{papr}$, where $\lambda$ = 0.01\\
\textbf{Step 6: Training} \\
Train the AE model with backpropagation to optimize $Loss_{total}$; repeat for $N$ iterations or until convergence. \\
\textbf{Step 7: Evaluation} \\
Calculate BER and CCDF for PAPR performance, and output the reconstructed sequence $\hat{\mathbf{x}}$ with optimized PAPR. \\
\end{algorithm}

\section{Simulation Results}
\begin{figure}[htbp]
    \centering
    \includegraphics[width=0.9\linewidth]{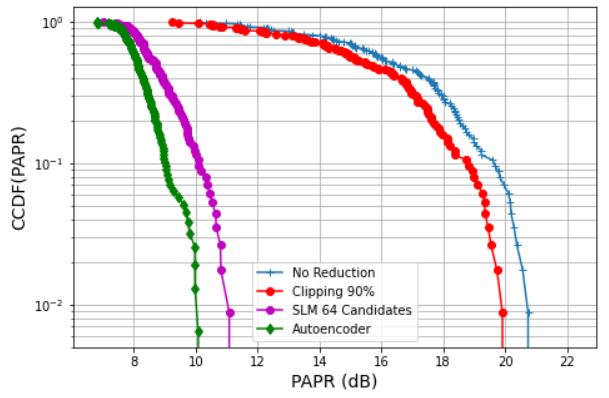}
    \caption{CCDF of OFDM signals}
    \label{Res1}
\end{figure}

In this section, we evaluate the performance of the proposed AE-based PAPR reduction method within a CO-OFDM system. The simulation results compare the effectiveness of AE with traditional methods, such as clipping and SLM, in terms of both PAPR reduction and BER performance under varying channel conditions. By analyzing CCDF and BER versus OSNR, we demonstrate the ability of the AE model to enhance the system resilience against noise and nonlinearities, ultimately achieving higher efficiency and reliability in optical communications. In this regard, Fig.~\ref{Res1} shows the complementary cumulative distribution function (CCDF) of the transmitted OFDM signals. The CCDF is a measure of the probability that the PAPR of the OFDM signals exceeds a certain value. The results demonstrate that AE outperforms both clipping and SLM in terms of PAPR reduction. Using AE, the effectiveness of PAPR reduction is even better when using the SLM reduction technique utilizing 64 phase sequence candidates. It achieves a PAPR reduction of more than 10 dB compared to no reduction and 1 dB over the SLM. This implies the effectiveness of the proposed AE in encoding the signal in a way that minimizes PAPR.    
\begin{figure}[htbp]
    \centering
    \includegraphics[width=0.9\linewidth]{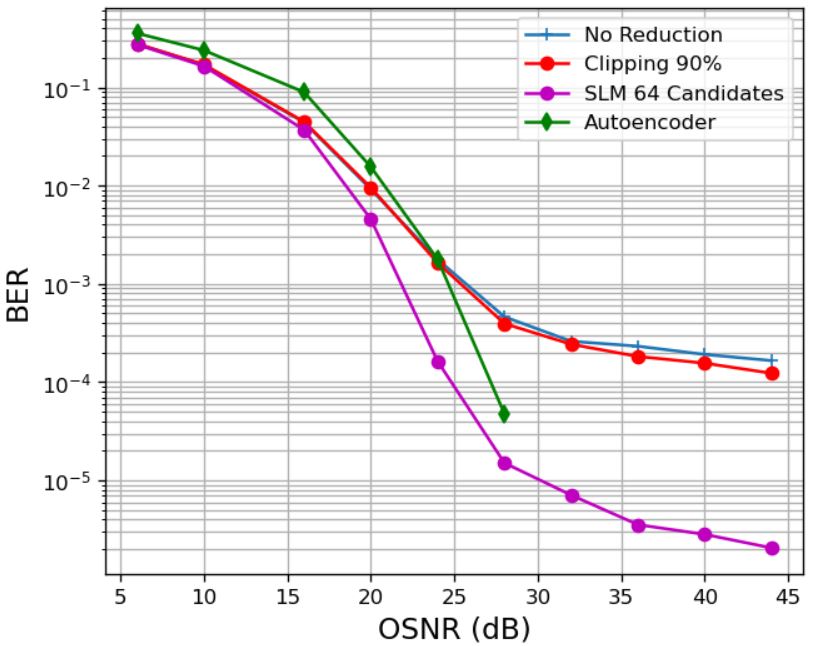}
    \caption{AE performance versus various PAPR reduction approaches in relation to BER and OSNR}
    \label{Res2}
\end{figure}
Fig.~\ref{Res2} illustrates that using AE, which has been trained with a noise standard deviation of 0.35 ($\sigma$), allows error-free transmission at an optical signal-to-noise ratio (OSNR) greater than approximately 28 dB. However, below 25 dB OSNR, no significant improvement is observed with the AE model, due to a significantly higher noise level. In this regime, noise dominates the signal, limiting the ability of AE to enhance BER despite being trained for noise handling. When noise levels are too high, the AE cannot significantly outperform traditional methods such as SLM or clipping. In contrast, systems not utilizing AE are incapable of attaining error-free transmission, leading to the manifestation of an error floor across all transmission methodologies (encompassing those without PAPR reduction, as well as those implementing the clipping and SLM techniques). For example, the transmission system with SLM employing 64 candidates achieves a BER of approximately \(2 \cdot 10^{-6}\) at 44 dB OSNR, while AE achieves error-free transmission starting from 32 dB OSNR. This shows that the system employing AE can achieve a remarkable improvement in BER, especially when the optical signal is recoverable in the presence of noise. This can be explained by the fact that the system is being optimized to efficiently handle such noise and nonlinearity effects through learning processes.

\begin{figure}[htbp]
    \centering
    \includegraphics[width=0.9\linewidth]{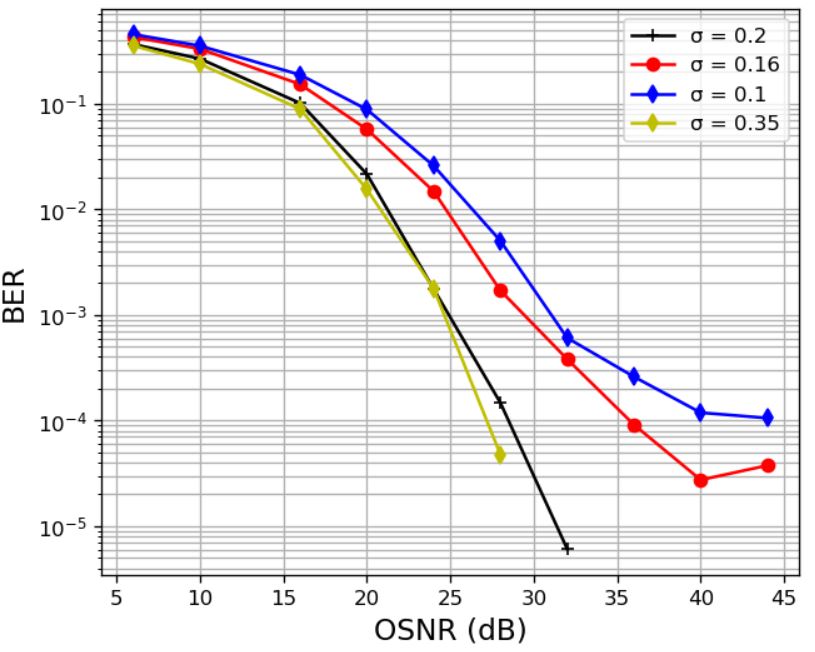}
    \caption{BER vs. OSNR: Autoendcoder learning with different noise standard deviation}
    \label{Res4}
\end{figure}

Various values of the noise standard deviation ($\sigma$), namely 0.1, 0.16, 0.2, and 0.35, were investigated to evaluate the performance of the AE transmission system. Fig.~\ref{Res4} illustrates that error-free transmission can be achieved by the transmission system when using $\sigma$ values of 0.2 and 0.35, at approximately 32 dB and 28 dB, respectively. In contrast, for noise standard deviation values of 0.1 and 0.16, the system does not achieve error-free transmission and instead experiences an error floor.

\section{Conclusions}
The proposed autoencoder-based model for PAPR reduction in CO-OFDM systems exhibits substantial enhancements in both PAPR mitigation and BER performance. The complementary CCDF plots show a notable decrease in the probability of high-PAPR events. The model not only reduces PAPR, but also leads to improved BER performance by performing AE training with a combined loss function that considers both PAPR and BER performance. In addition, the AE training model incorporates varying Gaussian noise standard deviations within the channel layer, enabling it to generate robust symbol sequences against noise. Compared to the SLM-based PAPR reduction model, the proposed model shows promising results, especially in terms of PAPR reduction and BER performance, making it a valuable approach for improving the performance of the optical OFDM system. In addition, it simplifies the process by eliminating the need for additional IFFT operations and complex search algorithms. Furthermore, our approach does not require any side information, whereas the SLM relies on knowledge of the PAPR distribution. The proposed AE-based model has significant potential to revolutionize next-generation optical communication systems. By efficiently reducing PAPR, it can ensure reliable data transmission over optical channels, even in the presence of various impairments.

\bibliographystyle{IEEEtran}
\bibliography{references}

\end{document}